\documentclass[12pt]{article}
\usepackage{amsmath,amssymb,amsthm,amsxtra,overpic,bbm,bm,epsfig,subfigure}
\usepackage{mathrsfs}
\usepackage{graphicx}
\usepackage{epstopdf}
\usepackage{color}
\usepackage{comment}
\usepackage{float}
\usepackage{cite}
\usepackage{slashed,stmaryrd}

\begin{document}

\vspace{0.2cm}
\begin{center}
{\Large\bf Sterile neutrinos as a possible explanation for the upward air shower events at ANITA}
\end{center}
\vspace{0.2cm}

\begin{center}
{\bf Guo-yuan Huang$^{a, b}$} \footnote{E-mail: huanggy@ihep.ac.cn}
\\
\vspace{0.2cm}
{\small $^a$Institute of High Energy Physics, Chinese Academy of
Sciences, Beijing 100049, China \\
$^b$School of Physical Sciences, University of Chinese Academy of Sciences, Beijing 100049, China}
\end{center}

\vspace{1.5cm}

\begin{abstract}
The ANITA experiment has observed two unusual upgoing air shower events which are consistent with the $\tau$-lepton decay origin. However, these events are in contradiction with the standard neutrino-matter interaction models as well as the $\rm EeV$ diffuse neutrino flux limits set by the IceCube and the cosmic ray facilities like AUGER. In this paper, we have reinvestigated the possibility of using sterile neutrino hypothesis to explain the ANITA anomalous events. The diffuse flux of the sterile neutrinos is less constrained by the IceCube and AUGER experiments due to the small active-sterile mixing suppression. The quantum decoherence effect should be included for describing the neutrino flux propagating in the Earth matter, because the interactions between neutrinos and the Earth matter are very strong at the EeV scale. After several experimental approximations, we show that the ANITA anomaly itself is able to be explained by the sterile neutrino origin, but we also predict that the IceCube observatory should have more events than ANITA. It makes the sterile neutrino origin very unlikely to account for both of them simultaneously. A more solid conclusion can be drawn by the dedicated ANITA signal simulations.
\end{abstract}

\begin{flushleft}
\hspace{0.8cm} PACS number(s): 14.60.St, 14.60.Pq, 98.70.Sa
\end{flushleft}

\newpage
\section{Introduction}
The cosmic rays with energy around the Greisen-Zatsepin-Kuzmin (GZK) cutoff ($\sim 50~{\rm EeV}$) would be strongly suppressed due to the interactions with the cosmic microwave background \cite{Greisen:1966jv, Zatsepin:1966jv}, i.e., $p+\gamma_{\rm CMB} \rightarrow p~({\rm or}~n)+{\rm \bf n} \pi$, $p+\gamma_{\rm CMB} \rightarrow \Delta^{+}(1232) \rightarrow p+ \pi^{0}~({\rm or}~n+\pi^{+})$, where ${\rm \bf n}$ is the total number of the produced $\pi$'s. The ultrahigh energy (UHE) neutrinos at the scale of EeV could be copiously produced by the subsequent decays of secondary charged pions and neutrons \cite{Beresinsky:1969qj}. The EeV neutrino remains to be detected, and the Antarctic Impulsive Transient Antenna (ANITA) experiment \cite{Gorham:2008dv} is dedicated to the detection of these cosmogenic neutrinos. 

In 2016, the ANITA experiment has reported one unusual steeply upward-pointing cosmic ray event 3985267 with shower energy around $0.6~{\rm EeV}$ during the ANITA-I flight \cite{Gorham:2016zah}. This shower event has the characteristics of the decay of a $\tau$-lepton, which is emerging from the surface of the ice with the zenith angle around $63^{\circ}$ \footnote{The reported emergence angle of the event 3985267 is $\sim 27^{\circ}$ below the horizontal, the corresponding zenith angle is thus $\sim 63^{\circ}$. One should be aware that the ANITA horizon is around $6^{\circ}$ below it's horizontal because of it's altitude.}, and the $\tau$-lepton should be interpreted as the product of a parent $\nu_{\tau}$ by the charged-current (CC) interactions with the Earth matter. However, such a hypothesis is strongly disfavored due to that the Earth CC attenuation coefficient is ~$4\times 10^{-6}$ for the neutrinos coming from such a steep angle \cite{Gorham:2016zah}. The associated event number around $E_{\nu} \sim 1~{\rm EeV}$ is negligible after adopting the IceCube bound on the diffuse EeV neutrinos \cite{Aartsen:2017mau}, approximately $E^2_{\nu} {\mathrm{d}\Phi_{\nu}}/{\mathrm{d} E_{\nu}} \lesssim 2\times 10^{-8}~{\rm GeV\cdot cm^{-2}s^{-1}sr^{-1}}$. In addition, there should be more Earth-skimming events than the steep events.  The situation is worse after the ANITA detector observed the second such air shower event 15717147 with energy around $0.56~{\rm EeV}$ at a steeper zenith angle $\sim 55^{\circ}$ during the ANITA-III flight \cite{Gorham:2018ydl}. Possible explanations for the anomalous events including the large transient point-source flux \cite{Gorham:2018ydl}, the transition radiation of the Earth-skimming neutrinos \cite{Motloch:2016yic}, the sterile neutrino origin \cite{Cherry:2018rxj}, and the decay of the quasi-stable dark matter in the Earth's core \cite{Anchordoqui:2018ucj} have been investigated in the literature.

After the report of the first anomalous event, Ref. \cite{Cherry:2018rxj} has proposed that the sterile neutrinos could be the origin of such an event. The sterile neutrinos are well motivated by several particle physics issues and experimental anomalies. The heavy sterile neutrinos can explain the mass generation of light neutrinos through the seesaw mechanism \cite{Minkowski:1977sc,Yanagida:1979as,Glashow:1979nm,GellMann:1980vs}, the sterile neutrino in the $\rm keV$ mass range is a good candidate of the warm dark matter \cite{Adhikari:2016bei}, and the anomalies of the short baseline experiments like LSND and MiniBOONE, the Gallium source experiments as well as the reactor neutrino experiments hint at the existence of the $\rm eV$-scale sterile neutrinos \cite{Aguilar:2001ty,AguilarArevalo:2008rc,Giunti:2010zu,Mention:2011rk,Abdurashitov:2009tn,Kaether:2010ag}. To explain the ANITA anomalous events, we need a strong sterile neutrino flux. The sources of the flux could be the superheavy dark matter decays \cite{Aartsen:2018mxl,Ema:2013nda,Esmaili:2013gha,Feldstein:2013kka,Ko:2015nma,Kachelriess:2008bk}, the topological defects \cite{Kachelriess:2008bk} or some exotic interaction \cite{Cherry:2014xra,Cherry:2016jol,Ahlgren:2013wba,Huang:2017egl,Farzan:2018gtr,Jeong:2018yts,Berryman:2018jxt} which converts active neutrinos into sterile neutrinos during their propagation. When the sterile flux goes through the Earth, they will experience a suppressed cross section due to the small active-sterile mixing. In such a way, the neutrinos can make their way to the thin interaction region below the ANITA detector, finally producing the $\tau$-lepton by the CC interaction with the ice, water or rock inside the interaction region. However, according to the analysis of Ref. \cite{Cherry:2018rxj}, the sterile origin is in mild tension with the steep emergence angle, e.g. only $10\%$ of the events are expected to emerge with the zenith angle smaller than $63^{\circ}$ for an active-sterile mixing angle $\theta =0.1$. Obviously, the second event 15717147 reported later \cite{Gorham:2018ydl} with zenith angle $55^{\circ}$ sharpens the tension. 

In our work, the sterile neutrino origin is reexamined, and we mainly have two treatments different from Ref. \cite{Cherry:2018rxj} which are addressed as follows:
\begin{itemize}
\item The neutrinos will lose coherence when strongly interacting with the ambient matter. After the sterile neutrino mass eigenstate flux entering the Earth, the matter will frequently measure the neutrino states such that the survived flux will collapse into the sterile state $\nu_{\rm s}$.
\item Because a positive detection is made only when the payload of ANITA is covered by the induced impulse cones with angle around $1^{\circ}$ \cite{Gorham:2018ydl}, only a very small fraction of the  plane flux from each direction can be detected. Thus the effective area $A_{\rm eff}{(\Omega)}$ should be much smaller than the expectation of Ref. \cite{Cherry:2018rxj}.
\end{itemize}
This work is organized as follows. In section 2, we investigate the evolution of the sterile neutrinos propagating in matter with the decoherence effect included, then the angular dependence is studied. In section 3, we give our predictions of the ANITA events for different sterile neutrino parameters based on several assumptions and approximations of the experimental setup. In section 4, we make our conclusion.
\section{Propagation of Sterile Neutrinos}
Regardless of the EeV sterile neutrino sources, the sterile neutrinos will lose their coherence after a long distance of galactic travelling. They will form the mass eigenstate fluxes $\nu_4$ and $\nu_1$ with fractions of $\cos^2{\theta}$ and $\sin^2{\theta}$, respectively, where $\theta$ is the active-sterile mixing angle, $\nu_1$ harmlessly represents the three active neutrino mass eigenstates, and $\nu_4$ can carry the mass at keV scale. When the $\nu_4$ flux propagates into the Earth, the Earth matter will frequently interact with, or equivalently measure, the neutrino's flavor. The $\nu_4$-state is the superposition of the active and sterile components, i.e. $\nu_4 = \sin{\theta}~\nu_{\rm a}+\cos{\theta}~\nu_{\rm s}$, and only the active component $\nu_{\rm a}$ \footnote{Since the overlaps of $\nu_{e}$-$\nu_{4}$ and $\nu_{\mu}$-$\nu_{4}$ are strongly constrained, the left available active component would be $\nu_{\tau}$, actually $\nu_{\tau}$ is what we are interested in. See \cite{Blennow:2018hto} and the references therein for recent constraints on the sterile neutrinos mixing.} is able to collide with the ambient matter through the CC or neutral current (NC) interactions. The matter will serve as the quantum discriminator which can resolve the mixing, making the $\nu_4$-state to collapse into either $\nu_{\rm a}$-state or $\nu_{\rm s}$-state, we refer the reader to \cite{Harris:1980zi,Stodolsky:1986dx,Raffelt:1992uj} for more details. To properly take the decoherence effect into account, we adopt the following evolution equation
\begin{eqnarray} \label{eq:Evolution}
i\frac{\mathrm{d}}{\mathrm{d} t} \left(\begin{matrix} c_{\tau} \\ c_{\rm s} \end{matrix}\right) = \frac{1}{2E_{\nu}}\left[ U\left(\begin{matrix} m^2_1 & 0 \\ 0 & m^2_4 \end{matrix}\right) U^{\dagger} + \left(\begin{matrix} A_{\rm NC} & 0 \\ 0 & 0 \end{matrix}\right) - i \left(\begin{matrix}  E_{\nu}/L_{\rm atten} & 0 \\ 0 & 0 \end{matrix}\right)\right]\left(\begin{matrix} c_{\tau} \\ c_{\rm s} \end{matrix}\right) 
\end{eqnarray}
for $\nu_4(t) = c_{\tau} \nu_{\tau} + c_{\rm s} \nu_{\rm s}$, where $U$ is the $2\times 2$ active-sterile mixing matrix with the mixing angle $\theta$, $m_1$ is the averaged active neutrino mass which is negligible compared with $m_4$ at keV scale, $A_{\rm NC} = -G_{\rm F}  (1- Y_e) n_{\rm N} / \sqrt{2}$ is due to the NC matter effect with $G_{\rm F}$ being the Fermi constant, $Y_e$ the fraction of electrons and $n_{\rm N}$ the nucleon number density of the matter, and $L_{\rm atten}$ is the local attenuation length of the neutrino. $L_{\rm atten}$ depends on the nucleon density and the neutrino energy through $L_{\rm atten} = [\sigma(E_{\nu}) n_{\rm N}]^{-1}$. The number density profile of the Earth can be found in the PREM model \cite{Dziewonski:1981xy}. The NC and CC cross sections are referred to \cite{Gandhi:1995tf}, and we note that both the CC and NC interactions contribute to the attenuation effect. We have neglected the regeneration effects for simplicity. The $\tau$-lepton produced by the CC can decay back to $\nu_{\tau}$. For the NC interaction, the produced neutrino carries averagely $80\%$ of the initial energy, but not removed from the flux. Thus our simulation will be more conservative than the realistic case. The initial conditions for the evolution read $c_{\tau} (0) = \sin{\theta}$, $c_{\rm s} (0) = \cos{\theta}$ before the $\nu_4$ flux entering the Earth. 

Before turning to numerical demonstration of the evolution, we can first have some analytical observations. 
\begin{figure}[t]
\centering
\includegraphics[scale=0.75]{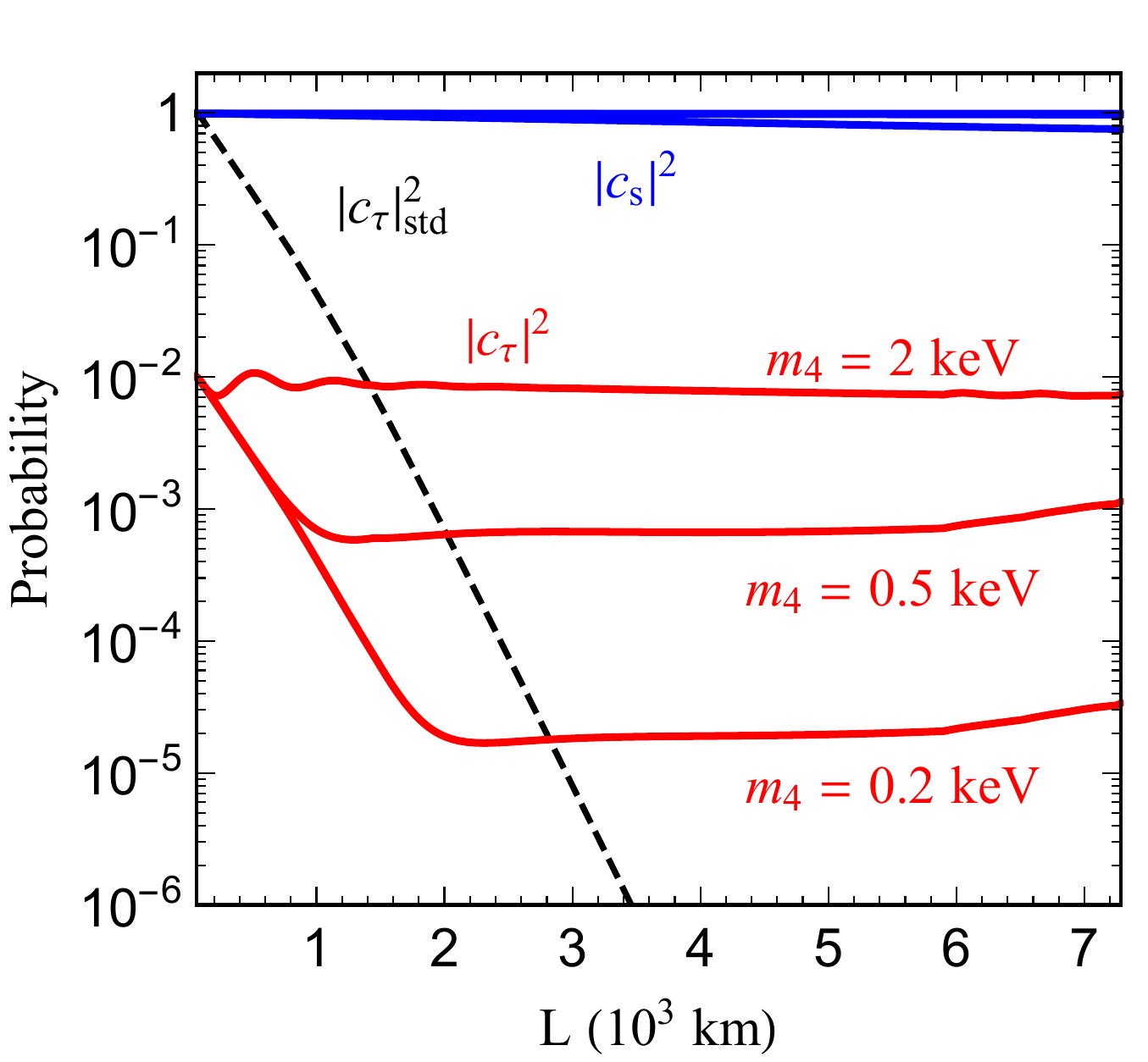}
\caption{The evolution of the EeV neutrino survival probability with respect to the travelling distance, corresponding to the ANITA event  15717147. The active-sterile mixing angle is chosen as $\theta=0.1$, and the mass of $\nu_4$ could be $2~{\rm keV}$, $0.5~{\rm keV}$, or $0.2~{\rm keV}$. The dashed curve shows the evolution of the standard $\nu_{\tau}$ flux. The solid blue curve stands for the survival probability of the sterile component, while the solid red curves stand for that of the $\nu_{\tau}$ component in the context of active-sterile mixing. }
\label{fig:Evolution}
\end{figure}
If we ignore the oscillation terms, i.e. the first two terms in the right-hand side of Eq.~({\ref{eq:Evolution}}), the evolution is trivial. The active and sterile components will evolve independently such that the active component is quickly absorbed by the Earth with only the unobservable sterile component left, and there will be null signal in the detector as in the standard case. However, the sterile and active neutrinos are actually mixed, and can oscillate from one to the other if the propagation length covers the oscillation length of $L_{\rm osc} \equiv 4\pi E_{\nu} / \Delta m^2_{41} \approx  2476~{\rm km}~[E_{\nu}/ (1~{\rm EeV})]~[(1~{\rm keV})^2/\Delta m^2_{41}]$. For the ANITA events 3985267 and 15717147 with emitting zenith angles of $63^{\circ}$ and $55^{\circ}$, the corresponding chord lengths are $5785~{\rm km}$ and $7309~{\rm km}$ respectively, assuming a spherical Earth shape. Therefore, it's quite evident that $\nu_4$ mass should be around the ${\rm keV}$ scale or even larger to convert the $\nu_{\rm s}$ flux into the $\nu_{\tau}$ flux when traversing the Earth. In such a way, the $\nu_{\tau}$ flux can be regenerated and survive the attenuation of the Earth. For the ${\rm keV}$-sterile neutrinos, the Mikheyev-Smirnov-Wolfenstein (MSW) resonance condition inside the Earth can be fulfilled. However, the associated total flavor conversion effect at the resonance point as in the discussion of solar neutrino problem is not available here. The matter term in the Earth mantle for the EeV neutrino reads $A_{\rm NC} \approx 0.1~{\rm keV}^2$. To satisfy the resonance condition $A_{\rm NC}= \Delta m_{41}^2 \cos{\theta}$ for the antineutrinos with mixing angle $\theta$, the $m_4$ should be around $0.3~{\rm keV}~[\sqrt{\cos\theta}]^{-1}$. The effective mass-squared difference when the resonance is achieved reads $\Delta\tilde{m}^2_{41}= \Delta m^2_{41} \sin{2\theta} \approx 0.2\sin{\theta}~{\rm keV}^{2}$, which corresponds to an oscillation length of $[12370/ \sin{\theta}]~{\rm km}\sim R/\sin{\theta}$ with $R = 12742~{\rm km} $ being the diameter of the Earth. Note that the density of the Earth changes very rapidly, by $20\%$ for the Earth mantle of $2000~{\rm km}$. It's very unlikely for those neutrinos to stay around the resonance while developing the phase. Our numerical calculation has included the matter effect without making any approximations.

In Fig.~1, we show the evolution of the EeV neutrino fluxes with respect to the travelling distance in the Earth. With the zenith angle of $\theta_z = 55^{\circ}$, the corresponding chord length through the Earth matter is around $7309~{\rm km}$. The dashed curve demonstrates the attenuation effect for the standard active neutrinos. Ref. \cite{Gorham:2016zah} has given the Earth attenuation factor as $4\times 10^{-6}$ for the event 3985267. Our numerical results of the attenuation factor are $1.2\times 10^{-9}$ for the event 3985267, and $1.4 \times 10^{-13}$ for the event 15717147, with both CC and NC interactions taken into account \footnote{As has been mentioned before, Ref. \cite{Gorham:2016zah} has only considered the CC interaction in their estimation of the attenuation length. We have included the NC interaction for the conservative purpose. Note that the actual attenuation factor should be larger after the regeneration effect included.}. The solid blue curve represents the survival probability of the sterile neutrino component $|c_{\rm s}|^2$, and it stays almost around one during the propagation. The solid red curves show the evolutions of the $\nu_{\tau}$ component in $\nu_4(t)$ with the masses of $2~{\rm keV}$, $0.5~{\rm keV}$, and $0.2~{\rm keV}$, respectively. As has been expected before, the survival probability of $\nu_{\tau}$ drops with the decreasing $\nu_4$-mass because of the increasing oscillation length. For $m_{\rm 4}=2~{\rm keV}$, the survival probability of $\nu_{\tau}$ fluctuates around 0.01 due to the continuous regeneration from $\nu_{\rm s}$ flux, just as the Earth being transparent.

Both of the two ANITA events have energys around $0.6~{\rm EeV}$, i.e., $E_{3985267}=0.6\pm 0.4~{\rm EeV}$, $E_{15717147}=0.56^{+0.3}_{-0.2}~{\rm EeV}$. 
\begin{figure}[t]
\centering
\includegraphics[scale=0.75]{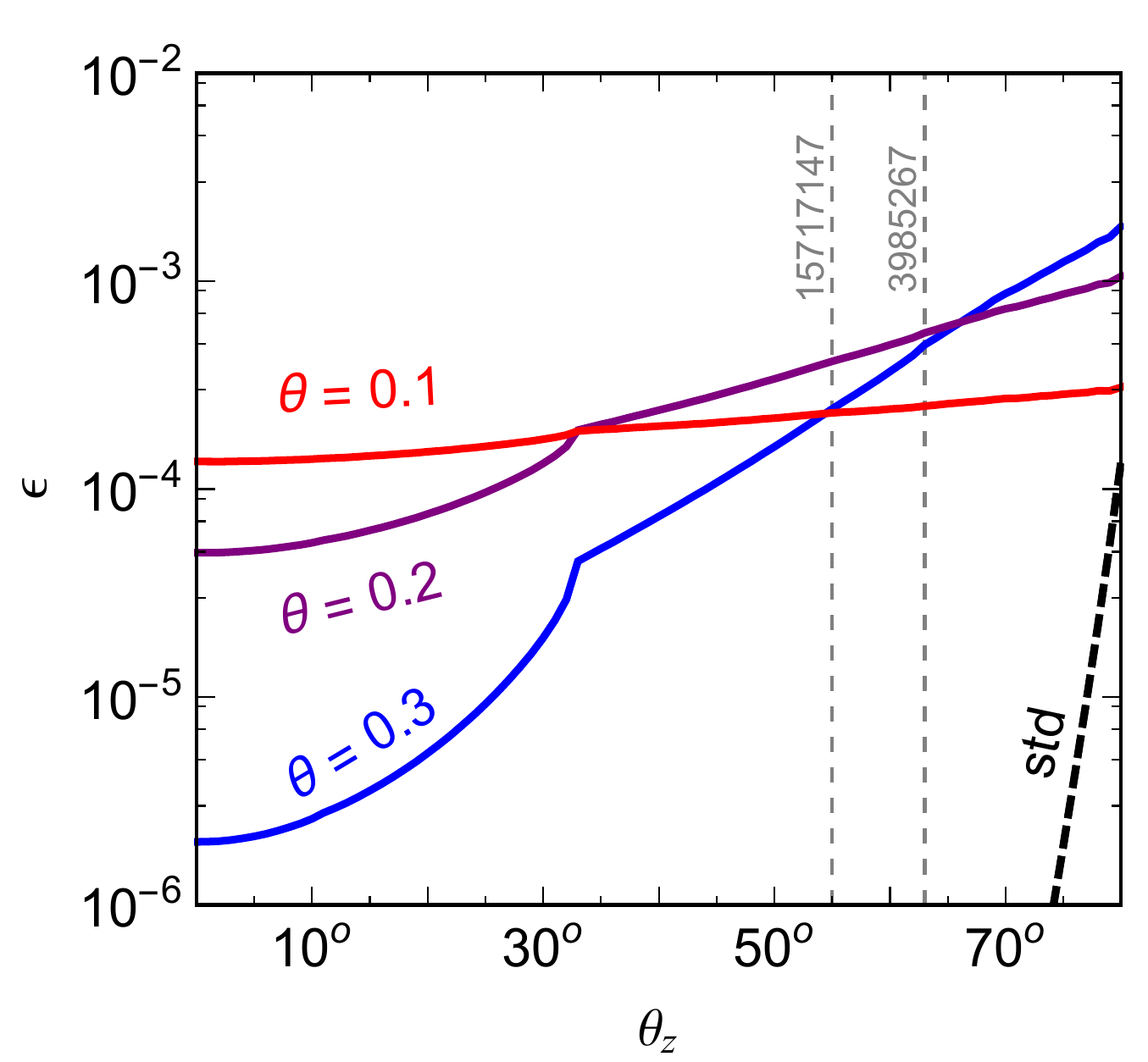}
\caption{The $\tau$-lepton transforming efficiency $\epsilon$ with respect to the emergency zenith angle. The sterile neutrino mass is fixed as $m_4=2~{\rm keV}$. The red, purple and blue curves stand for the cases with mixing angles of $\theta=0.1$, $\theta=0.2$ and $\theta=0.3$, respectively. The dashed curve in the bottom-right corner is just the efficiency of the pure active neutrino flux $\nu_{\tau}$. The two vertical lines correspond to the emergence angles of the two ANITA events $15717147$ and $3985267$. Note that $\theta_z \gtrsim 80^{\circ}$ is nearly above the ANITA horizon. The knee around $\theta_z = 30^{\circ}$ is due to the large density jump between the Earth outer core and mantle. The energy loss of $\tau$-leptons are simulated with the ASW model \cite{Armesto:2004ud,Alvarez-Muniz:2017mpk}.}
\label{fig:ea}
\end{figure}
To simplify our calculation, we assume that the initial $\nu_4$ flux has almost monochromatic energy around $1~{\rm EeV}$. After these neutrinos entering the Earth, they can propagate almost freely to the other side of the Earth just as the $m_{\rm 4}=2~{\rm keV}$ case in Fig. 1. With a $\nu_{\tau}$ residue in the interaction region with depth around tens of kilometer below the ANITA balloon, the flux is eventually transformed into observable $\tau$-lepton flux by the CC interaction. We define the efficiency of the initial $\nu_4$ particles transformed into $\tau$-lepton as
\begin{eqnarray} \label{eq:efficiency}
\epsilon(\Omega) = \frac{ \mathrm{d} \Phi_{\tau}(E_{\rm min}, E_{\rm max})/ \mathrm{d} \Omega}{\mathrm{d} \Phi_{\nu_{4}}(E_{0}) / \mathrm{d} \Omega},
\end{eqnarray}
where $E_0=1~{\rm EeV}$ is the initial neutrino energy, $\Phi_{\nu_4}$ stands for the isotropic flux of neutrinos, $\Phi_{\tau}(E_{\rm min}, E_{\rm max})$ is the produced $\tau$-lepton flux in the energy range of $[E_{\rm min}, E_{\rm max}]$ when they arrive at the Antarctic surface. We set $E_{\rm min} = 0.2~{\rm EeV}$ and $E_{\rm max} = 1~{\rm EeV}$ for our calculation, so that the two observed ANITA events are covered by $[E_{\rm min}, E_{\rm max}]$. The transforming efficiency $\epsilon(\Omega)$ measures the fraction of neutrinos being converted into the observable $\tau$-leptons during their way to ANITA, it should be noted that $\epsilon(\Omega)$ is direction-dependent. In Fig. 2, we show the angular dependence of the $\epsilon(\Omega)$ for the cases with $m_{\rm 4} = 2~{\rm keV}$.
The dashed curve in the bottom-right corner shows the efficiency in the standard scenario. The $\tau$-leptons induced by the standard isotropic active neutrino flux should be concentrated around the large zenith angles, therefore the Earth-skimming shower should dominate the events as has been expected. The red curve with active-sterile mixing angle of $\theta=0.1$ is almost uniformly distributed for the entire zenith angle range. The efficiency is around $10^{-4}$, which means that no matter which angle the neutrino flux comes from, there will always be one observable $\tau$-lepton emitted after $10^4$ $\nu_4$ neutrinos entering the Earth. The Earth-skimming events don't have much advantage over the steep events in this case. However, as the active-sterile mixing angle increases, the sterile neutrino would become not so sterile due to the large mixing with the active neutrino. The efficiency tends to converge into the standard case.

\section{ANITA Events Estimation}
Due to the small Cherenkov ring angle of the EeV $\tau$-decay shower, only a very small fraction of the $\tau$-lepton flux obtained in the last section can be captured by the antennas of ANITA. \begin{figure}[t]
\centering
\includegraphics[scale=0.75]{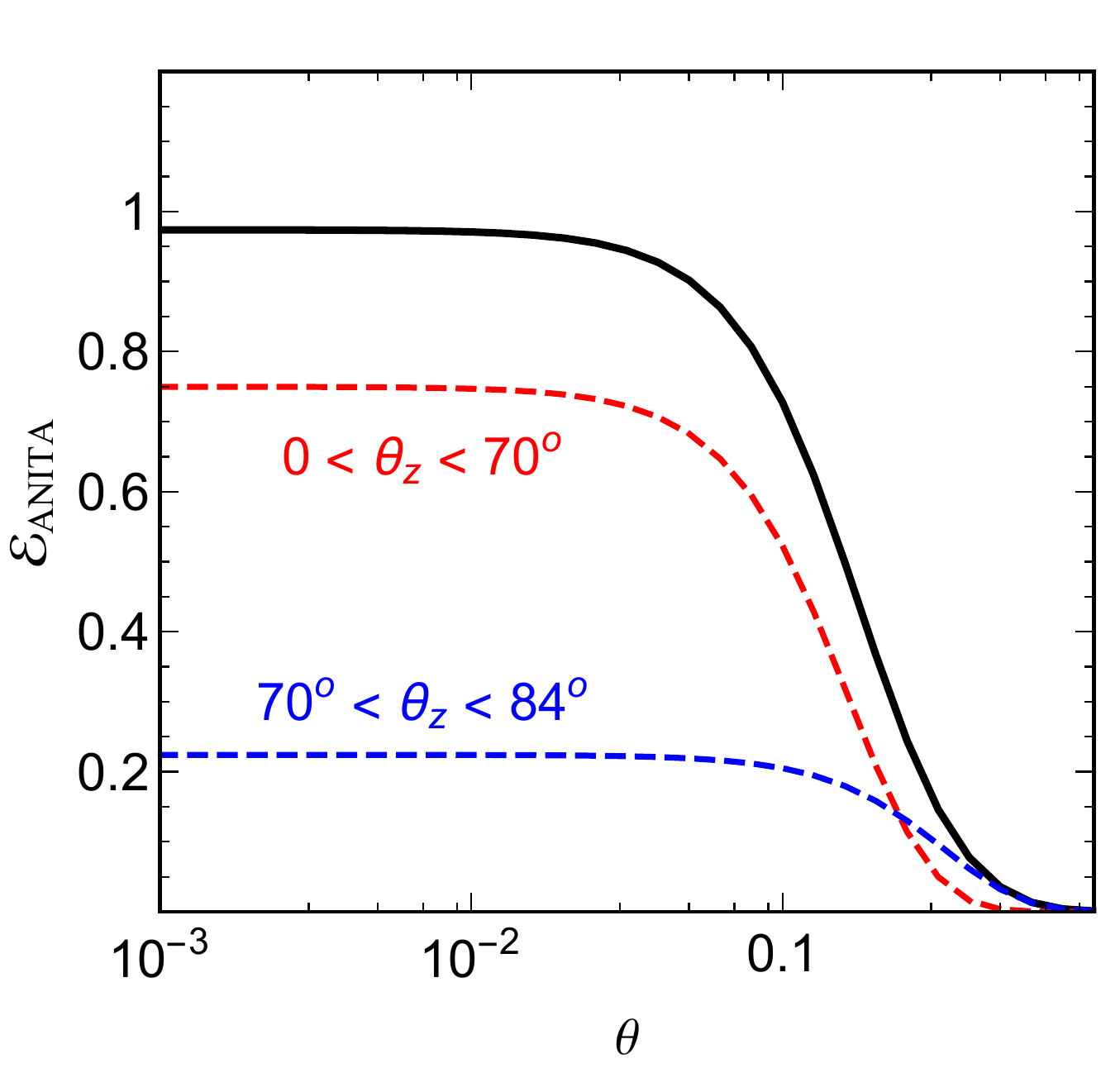}
\caption{The event number of ANITA for three months of exposure. For each mixing angle $\theta$, the corresponding IceCube bound on $\nu_4$ flux is saturated. The black solid curve is the event number for the total sky. The red (blue) dashed curve stands for the shower events emitted from $> 20^{\circ}~(< 20^{\circ})$ below the horizontal. $\theta_z = 84^{\circ}$ corresponds to the zenith angle of the ANITA horizon.}
\label{fig:events}
\end{figure}
The angle of the light cone should be around $1^{\circ}$ \cite{Gorham:2018ydl}, so we expect the geometric area of ANITA should be about $A_{\rm gm} \approx 2\pi (D \times 1^{\circ})^2$ with $D$ being the distance from the ANITA payload to the initial point of the shower. For event $15717147$ with $\theta_z = 55^{\circ}$, we obtain the geometric area as $\sim 7.5~{\rm km}^2$, slightly larger than the estimation of the ANITA group $\sim 4~{\rm km}^2$ \cite{Gorham:2018ydl}. The realistic geometric area estimation requires the dedicated Monte Carlo simulation, and we simply assume a constant geometric area of $4~{\rm km}^2$ for all emergence angles to proceed our estimation. Simulations show that the $\tau$-lepton decay shower from larger zenith angle would have smaller impulse power \cite{Romero-Wolf:2017}, thus the steep shower seems more likely to be detected by ANITA than the Earth-skimming shower in the realistic case. 
The flux of the EeV $\nu_4$ is bounded by the IceCube observation as ${\mathrm{d}\Phi_{\nu_4}}/{\mathrm{d} \Omega} \lesssim 2\times 10^{-15}~[0.1/\sin{\theta}]^2~{\rm  cm^{-2}s^{-1}sr^{-1}}$, note that the flux limit is relaxed by the mixing angle suppression. The final event is obtained as
\begin{eqnarray} \label{eq:event}
\mathscr{E}_{\rm ANITA} = \int \mathrm{d}\Omega \cdot \frac{\mathrm{d}\Phi_{\nu_4}}{\mathrm{d} \Omega}\times \epsilon(\Omega) \times A^{\rm ANITA}_{\rm gm}(\Omega) \times T_{\rm ANITA}\approx 0.9,
\end{eqnarray}
where $A^{\rm ANITA}_{\rm gm}(\Omega)$ is simply fixed to $4~{\rm km}^2$ as mentioned before, $T_{\rm ANITA}$ is the three months of exposure for ANITA, $\epsilon(\Omega)$ is obtained with $m_{\rm 4} = 2~{\rm keV},~\theta = 0.1$ in the last section, the flux takes the saturated value of the IceCube bound ${\mathrm{d}\Phi_{\nu_4}}/{\mathrm{d} \Omega} = 2\times 10^{-15}~[0.1/\sin{\theta}]^2~{\rm  cm^{-2}s^{-1}sr^{-1}}$. One can identify the effective area as $A_{\rm eff} (\Omega)= \epsilon(\Omega)A^{\rm ANITA}_{\rm gm}(\Omega) \approx 10^7~{\rm cm}^2$, much smaller than the estimation $A_{\rm eff} \approx 10^{11}~{\rm cm}^2$ of \cite{Cherry:2018rxj}. Event number of 0.9 is obviously consistent with the ANITA observation.
Let's check the situation for other experiments. The IceCube has a geometric area around $1~{\rm km}^2$ for the through-going track events, therefore we can estimate the event number of IceCube with six years observation as $\mathscr{E}_{\rm IC} = 6$, using $A^{\rm IC}_{\rm gm}(\Omega) \approx 1~{\rm km}^2$, $T_{\rm IC} \approx 6~{\rm years}$. As has been pointed out in \cite{Anchordoqui:2018ucj,Kistler:2016ask}, IceCube might already have observed one such $\tau$-track event with energy $\gtrsim 0.1~{\rm EeV}$ and emergence angle of $11.5^{\circ}$ below the horizon. The deposited energy of this track is $(2.6\pm 0.3)~{\rm PeV}$ \cite{Aartsen:2016xlq}, implying a $\mu$-lepton track with energy $\gtrsim 10~{\rm PeV}$ or a $\tau$-lepton track with energy $\gtrsim 0.1~{\rm EeV}$. No matter whether this event is an $\rm EeV$ $\tau$-lepton track or not, the IceCube observation is in considerable tension with ANITA under the sterile hypothesis. The non-observation of EeV neutrino events at AUGER shouldn't be a problem with viewing angle of only a few degrees below the horizon for the Earth-skimming events \cite{Aab:2015kma}, while the produced $\tau$-lepton flux can be almost uniformly distributed as in Fig. 2. 

In Fig. 3, we plot the relation between the ANITA's three months of event number and the active-sterile mixing angles. The mass of $\nu_4$ is chosen to be $2~{\rm keV}$, and the results don't differ much for $m_4 > 2~{\rm keV}$. The solid black curve is the total event number. The red dashed curve is the events within the zenith angle range of $[0^{\circ},70^{\circ}]$, which is larger than the blue one, i.e., the events within the zenith angle range of $[70^{\circ},84^{\circ}]$ for $\theta \lesssim 0.1$. Some comments on the figure are in order:
\begin{itemize}
\item Note that the event number curves in Fig. 3 can be extrapolated towards very small mixing angles, but the associated sterile neutrino flux is also required to be stronger. The flux should stay around the IceCube upper limit ${\mathrm{d}\Phi_{\nu_4}}/{\mathrm{d} \Omega} \approx 2\times 10^{-15}~[0.1/\sin{\theta}]^2~{\rm  cm^{-2}s^{-1}sr^{-1}}$.
\item The observations of ANITA and IceCube can't be well fitted at the same time by tuning the flux and parameters of the sterile neutrino. Independent from the properties of the sterile neutrino, the $\tau$-lepton decay events of ANITA and the through-going $\tau$-lepton track events of IceCube always have the following relation:
\begin{eqnarray} \label{eq:AI}
\frac{\mathscr{E}_{\rm IC}}{\mathscr{E}_{\rm ANITA} } \approx \frac{A^{\rm IC}_{\rm gm}\times T_{\rm IC}}{A^{\rm ANITA}_{\rm gm} \times T_{\rm ANITA}} \approx 6.
\end{eqnarray}
The unlikelihood that ANITA has two events while IceCube has one or null event can reach $3{\rm \sigma}$. Note that this estimation can also be applied to the scenario in \cite{Anchordoqui:2018ucj}, where the quasi-stable dark matter decay inside the Earth is the origin of the ANITA anomalous events. 
\end{itemize}
A more reliable result depending on the dedicated Monte Carlo simulations of the ANITA experiment is beyond the scope of the present work.

\section{Conclusion}
In this work, we have reinvestigated the possibility of using sterile neutrino origin to explain the ANITA anomalous events. We find that the quantum decoherence effect is very important to account for the propagation behavior of EeV neutrino flux. The $\nu_{\tau}$ flux can be regenerated by the oscillation of $\nu_{\rm s}$-state during their propagation inside the Earth. For the sterile neutrinos with $m_{4} \gtrsim 1~{\rm keV}$, the Earth is almost transparent to the $\nu_{\tau}$ component in the $\nu_4$ flux. In this way, the neutrinos can losslessly reach the interaction volume below the ANITA payload with very steep angles. We have estimated the ANITA and IceCube event number, and find that averagely the ANITA experiment is able to observe one event during the three months of exposure, while the IceCube is supposed to detect six events for it's six years of data taking. To resolve the ANITA anomaly itself, the flux of $\nu_4$ should be around the upper IceCube limit ${\mathrm{d}\Phi_{\nu_4}}/{\mathrm{d} \Omega} \approx 2\times 10^{-15}~[0.1/\sin{\theta}]^2~{\rm  cm^{-2}s^{-1}sr^{-1}}$. We have scanned the whole sterile neutrino parameter space, and find that the requirements on the sterile neutrino parameters are $m_{4} \gtrsim 1~{\rm keV}$, $\theta \lesssim 0.1$. If the dark matter decays are the sources of the sterile neutrinos, the mixing angle shouldn't be too small, because it might be too difficult for the dark matter decays to produce so much strong neutrino flux that saturates the IceCube bound. In our framework, the predicted EeV $\tau$-lepton track event number of IceCube is always averagely six times of the $\tau$-lepton decay shower event number of ANITA, i.e., ${\mathscr{E}_{\rm IC}}/{\mathscr{E}_{\rm ANITA} } \approx 6$. This result is independent of the sterile neutrino parameters as well as whether the regeneration effect is included or not. Even though there is an ${\cal O}(0.1~{\rm EeV})$ track candidate for the IceCube observation, these two experiments together are in strong tension with the sterile neutrino explanation. However, we expect the dedicated ANITA simulations to draw a more solid conclusion.
\section*{Acknowledgements}
I am indebted to S. Zhou for suggesting this work and for many valuable discussions and suggestions. I am also grateful to N. Nath and Q.R. Liu for useful discussions. This work is supported by the National Natural Science Foundation of China under grant No. 11775232.


\end{document}